\documentclass[fleqn,10pt]{wlscirep}
\usepackage[utf8]{inputenc}
\usepackage[T1]{fontenc}
\title{Metasurfaces for Efficient Digital Noise Absorption}

\author[1]{Ryoya Aihara}
\author[1,2*]{Hiroki Wakatsuchi}
\affil[1]{Department of Electrical and Mechanical Engineering, Graduate School of Engineering, Nagoya Institute of Technology, Nagoya, Aichi, 466-8555, Japan}
\affil[2]{Precursory Research for Embryonic Science and Technology (PRESTO), Japan Science and Technology Agency (JST), Kawaguchi, Saitama, 332-0012, Japan}

\affil[*]{wakatsuchi.hiroki@nitech.ac.jp}

\begin{abstract}
We numerically demonstrate two types of metasurface absorbers to efficiently absorb digital signals. First, we show that the digital waveforms used in this study contain not only a fundamental wave but also nonnegligible harmonic waves, which limits the absorption performance of a conventional metasurface absorber operating in only a single, finite frequency band. The first type of the proposed absorbers is designed using two kinds of unit cells, each of which absorbs either a fundamental frequency or third harmonic of an incident digital waveform. This dual-band metasurface absorber exhibits absorption performance exceeding that of the conventional metasurface absorber and more strongly dissipates the energy of a digital waveform. In addition, the second type of absorber exploits the concept of nonlinear analogous circuits to convert an incoming wave to a different waveform, specifically, a triangular waveform that has a larger magnitude at a fundamental frequency. Therefore, the incoming waveform is more effectively absorbed by this waveform-conversion metasurface absorber as well. Although still there remain some issues to put these digital signal absorbers into practice, including experimental validation, our results contribute to mitigating electromagnetic interference issues caused by digital noise and realising physically smaller, lighter digital signal processing products for the next generation.
\end{abstract}
\begin{document}

\flushbottom
\maketitle

\thispagestyle{empty}

\section*{Introduction}

Wireless communication is important as our modern life is supported by many devices, such as mobile phones and Wi-Fi/Bluetooth devices. Due to a growing number of wireless communication devices, however, such devices receive not only necessary signals but also unnecessary electromagnetic (EM) noise, which lowers the quality of the communication environment and potentially leads to temporal malfunction or permanent damage, which is an issue well known in the field of EM compatibility \cite{CCemcBook}. This EM interference issue can be readily solved by using absorbing materials that effectively convert the EM energy of an incident wave to thermal energy \cite{salisbury1952absorbent, MunkBook, knott2006radar}. As a result, the magnitude of the reflected wave can be reduced to a negligible level for sensitive communication devices and electronics. Although classic absorbers tend to have strong absorption performance but relatively bulky dimensions, new types of absorbers based on subwavelength periodic structures, or so-called metasurfaces \cite{MTMbookEngheta, calozBook, smithDNG1D, sievenpiper2010, yu2011light, zhang2019high, yuan2020fully, yuan2020independent}, can be designed with ultra-thin thicknesses while still maintaining a strong absorption effect. In general, these metasurface-based absorbers exploit dielectric losses, ohmic losses or both to dissipate the energy of an incoming wave at a frequency of interest \cite{TaoInIEEEselectedTopic, watts2012metamaterial, li2017nonlinear, mtmAbsPRLpadilla, mtmAbsOEpadilla, ultraThinAbs, wakatsuchi2012performance}. In addition, metasurface-based absorbers can operate at multiple frequencies by including several unit cells, each of which resonates at a different frequency, addressing EM interference issues in multiple bands simultaneously \cite{dual1, dualBandAbsTao, My1stAbsPaper}. Moreover, different signals can be distinguished even at the same frequency by introducing nonlinear circuits (e.g., diodes and transistors) to metasurfaces, thus protecting electronic devices from strong EM noise while permitting the transmission of small wireless communication signals even at the same frequency \cite{aplNonlinearMetasurface, wakatsuchi2013waveform, wakatsuchi2015waveformSciRep, kim2016switchable, li2017high}. However, most metasurface absorbers reported to date focus essentially on absorbing EM waves at a single frequency or within a finite bandwidth. In contrast, modern electronic devices such as smartphones and augmented reality (AR)/virtual reality (VR) products tend to include integrated circuits (ICs) or central processing units (CPUs) that generate digital signals in physically limited spaces. An emerging issue here is that as opposed to the waveforms targeted by conventional metasurface absorbers (usually sinusoidal waveforms), digital waveforms contain not only a fundamental wave but also nonnegligible harmonic waves. Therefore, digital noise cannot be efficiently absorbed by conventional metasurface absorbers with a single operating frequency or a limited bandwidth. For this reason, this study presents metasurface absorbers dedicated to absorbing digital waveforms. In particular, we report and compare two types of digital-waveform absorbers, both based on metasurfaces but with/without nonlinear circuits. Potentially, our results contribute to suppressing digital noise more efficiently and designing the next generation of smaller, lighter digital signal processing devices.

\section*{Results}
\begin{figure}[tb]
\centering
\includegraphics[width=\linewidth]{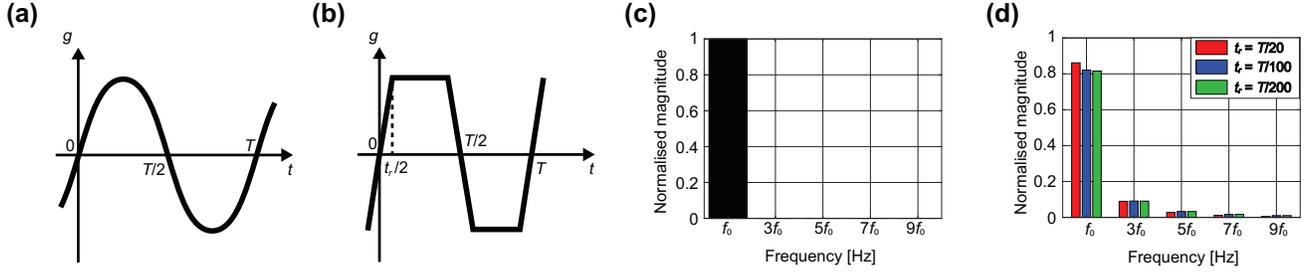}
\caption{(a) Sinusoidal waveform and (b) trapezoidal waveform. Magnitudes of the fundamental and harmonic waves in (c) the sinusoidal waveform and (d) the trapezoidal waveform. Here, $t_r$ represents the rise and fall times of the trapezoidal waveform. }
\label{fig:waveforms}
\end{figure}

\begin{table}[tb]
 \begin{center}
  \caption{Frequency components of the digital signal (i.e., trapezoidal waveform) with $t_r$ set to $T/20$.}
  \label{tab:traSpectrum}
  \begin{tabular}{lccccc} \hline \hline
Frequency & $f_0$ & $3f_0$ & $5f_0$ & $7f_0$ & $9f_0$  \\ \hline
Magnitude of power & 0.861 & 0.090 & 0.028 & 0.012 & 0.005  \\ \hline \hline
  \end{tabular}
 \end{center}
\end{table}

First, we show a difference between the general sinusoidal waveforms targeted by most conventional metasurface absorbers and the digital waveforms focused on in this study. The spectra of both waveforms can be calculated by using Fourier series, i.e.,
\begin{eqnarray}
\label{eq:1stFourier}
w\left(t\right) &=& \sum^{\infty}_{n=-\infty} C_n \mathrm{e}^{ j 2 \pi n f_0 t}, \\
\label{eq:1stFourierCoeff}
C_n &=& \frac{1}{T} \int^{T}_{0} w(t) \mathrm{e}^{-j 2 \pi n f_0 t} \: dt,
\end{eqnarray}
where $w$, $f_0$, $C$, $n$, $t$ and $T$ are the waveform, the fundamental frequency, the complex Fourier coefficient, an integer, the time and the period, respectively. In these equations, other variables, such as the phase delay, position in a particular coordinate system and waveform amplitude coefficient, are all omitted for the sake of simplicity. In the case of sinusoidal waves (Fig.\ \ref{fig:waveforms}a), eqs.\ (\ref{eq:1stFourier}) and (\ref{eq:1stFourierCoeff}) become
\begin{eqnarray}
C_n
&=& \begin{cases}
\displaystyle\frac{n}{2} &\left(n \colon 1 \; \rm{or} \; -1 \right) \\
0&\left(n \colon \rm{others} \right)
\end{cases},\\
w\left(t\right) &=& \sin 2 \pi f_0 t .
\end{eqnarray}
We assume that the digital waveforms used in this study (Fig.\ \ref{fig:waveforms}b) vary between a positive value and a negative value at a duty cycle of 50 \% for simplicity. In reality, however, such digital waveforms have a rise time and fall time, which means that the waveforms are closer to trapezoidal waves than square waves. Therefore, $w(t)$ and $C_n$ of digital waveforms are obtained by
\begin{eqnarray}
C_n
&=& \begin{cases}
\displaystyle \frac{\sin \displaystyle \frac{\pi n t_r}{T} }{\displaystyle \frac{\pi n t_r}{T}} \cdot \frac{\sin \displaystyle \displaystyle \frac{\pi n}{2} }{\displaystyle \frac{\pi n}{2}} &\left( n\colon \rm{odd} \right) \\
0 &\left( n\colon \rm{even} \right)
\end{cases},\\
w\left(t\right) &=& 2\sum^{\infty}_{n=1} \frac{\sin \displaystyle \frac{\pi n t_r}{T}}{\displaystyle \frac{\pi n t_r}{T}} \cdot \frac{\sin \displaystyle \frac{\pi n}{2} }{\displaystyle \frac{\pi n}{2}} \sin {2 \pi n f_0 t} ,
\label{eq:trpG}
\end{eqnarray}
where the rise and fall times are set to the same value $t_r$. Based on these equations, the magnitude of each frequency component of these waveforms is calculated as seen in Figs.\ \ref{fig:waveforms}c and d, where both waveforms are normalised to have the same power level in total. These figures analytically support that sine waves convey energy only through a fundamental wave, while digital waveforms (trapezoidal waves) convey energy not only through a fundamental wave but also through high harmonic waves (see also Table \ref{tab:traSpectrum} for specific values). In the following part of this study, we set $t_r$ to $T$/20.

\begin{figure}[tb]
\centering
\includegraphics[width=\linewidth]{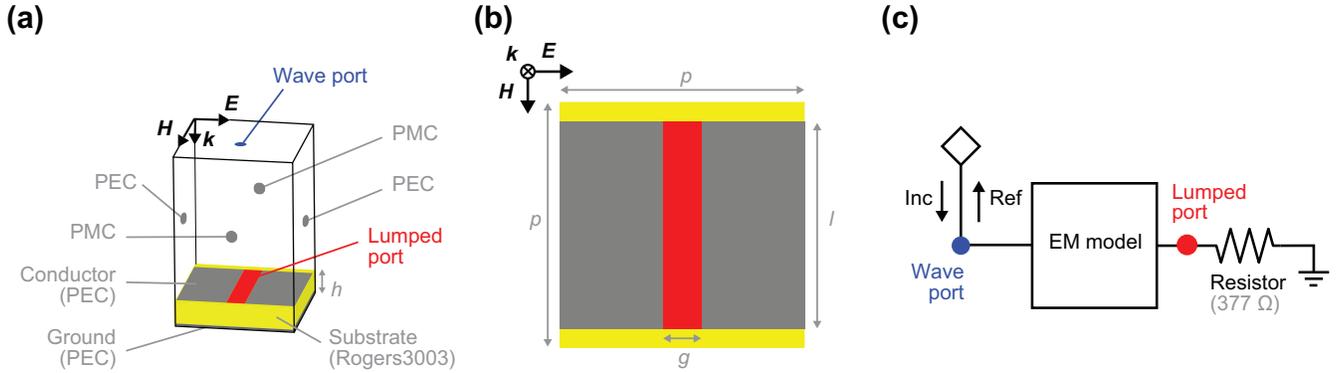}
\caption{(a, b) Model for EM simulations. (c) Schematic for circuit simulations. The design parameters are given in Table \ref{tab:conv}.}
\label{fig:coSim}
\end{figure}

First, we show the limited performance of a conventional metasurface absorber for digital waveforms. This structure was assumed to be composed of periodically deployed conducting square patches (perfect electric conductor, PEC), a dielectric substrate (Rogers3003 but without loss for the sake of simplicity) and a ground plane (see Figs.\ \ref{fig:coSim}a, b). In addition, conducting patches were connected by lumped resistors (377 $\Omega$) to dissipate the energy of an incoming wave. The design parameters of this structure are given in Table \ref{tab:conv}.

The absorption performance of this structure was numerically tested using a co-simulation method \cite{wakatsuchi2013waveform, wakatsuchi2019waveform} that integrated an EM solver with a circuit solver (ANSYS Electronics Desktop R18.1.0). This method is well known to have an advantage over a stand-alone EM simulation method from the viewpoint of simulation efficiency and speed, which contributed to facilitating the optimisation process of metasurface absorbers studied here, although EM field distributions cannot be visualised in this method \cite{wakatsuchi2015fieldVisualisation, TLMbook1, jin2015finite, taflove2005computational}. As illustrated in Fig.\ \ref{fig:coSim}, the first step of this method was modelling metasurface absorbers in the EM solver and performing EM simulations. However, all of the circuit components were replaced with lumped ports that were later connected to the actual circuit components in the circuit solver, where circuit simulations were performed as the second step to complete co-simulations. In the case of the conventional metasurface absorber, for example, a lumped resistor deployed in a gap between conducting patches was connected to an EM model via a lumped port in the circuit solver (see Fig.\ \ref{fig:coSim}c).

\begin{table}[tb]
 \begin{center}
  \caption{Design parameters used for the conventional metasurface absorber simulated in Figs.\ \ref{fig:coSim} and \ref{fig:conv}.}
  \label{tab:conv}
  \begin{tabular}{lcccc} \hline \hline
    Parameter & $p$ & $h$ & $g$ & $l$  \\ \hline 
    Value  & 17 & 3 & 1 & 16 \\ 
    Unit  & mm & mm & mm & mm \\ \hline \hline 
  \end{tabular}
 \end{center}
\end{table}

\begin{figure}[tb]
\centering
\includegraphics[width=0.7\linewidth]{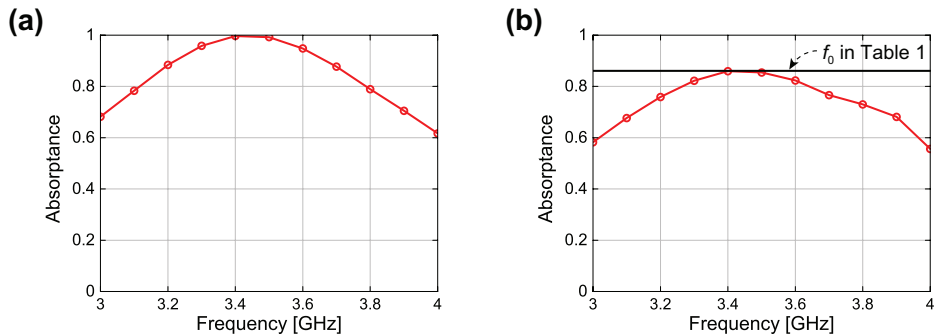}
\caption{Simulation absorptances of the conventional metasurface absorber for the (a) sinusoidal waveform and (b) digital waveform. }
\label{fig:conv}
\end{figure}

Under these circumstances, the metasurface absorber absorbed nearly 100 \% the energy of a sine wave near 3.4 GHz, where the wave impedance of the structure strongly matched that of free space (see Fig.\ \ref{fig:conv}a). However, when the incoming waveform changed to a trapezoidal form as a digital signal, the maximum absorptance reduced to approximately 86 \% (Fig.\ \ref{fig:conv}b), which was found not to be larger than the normalised energy of the fundamental wave of the trapezoidal waveform used (see the black horizontal line corresponding to $f_0$ of Table \ref{tab:traSpectrum}).

\begin{figure}[tb]
\centering
\includegraphics[width=\linewidth]{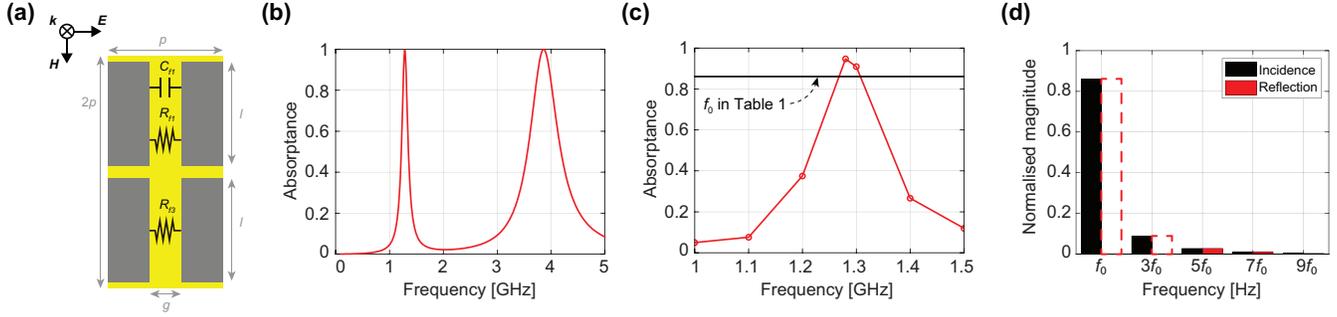}
\caption{(a) Super cell of a dual-band metasurface absorber comprising two types of unit cells operating at fundamental and third harmonic waves. The design parameters are given in Table \ref{tab:dual}. (b) Absorptance for the sinusoidal waveform. (c) Absorptance for the digital waveform. (d) Normalised magnitudes of both the incident and the reflected waves ($f_0$ set to 1.29 GHz). }
\label{fig:dual}
\end{figure}

Next, we present the first type of digital noise absorber using a linear metasurface (Fig.\ \ref{fig:dual}a, Table \ref{tab:dual}). This absorber had super cells composed of two kinds of unit cells, each of which operated at different frequencies, such as dual-band absorbers \cite{dual1, dualBandAbsTao, My1stAbsPaper}. Importantly, however, one of the operating frequencies was set to the fundamental wave of a digital signal, while another was set to its third harmonic. To readily achieve such a dual-band absorption mechanism, this metasurface absorber deployed one capacitor for every two gaps between conductors, which enabled us to adjust the two different operating frequencies even with the same physical dimension (Fig.\ \ref{fig:dual}a). The reason is that the resonant (operating) frequencies $f_0$s of metasurfaces fundamentally relate to their entire inductive component $L$ and capacitive component $C$ through $f_0=1/2\sqrt{LC}$, where $L$ and $C$ are determined by, for instance, the conducting geometries and dimensions of metasurfaces as well as lumped circuit components if used together \cite{baena2005equivalent, ZhouCWeq, MyCWeqCircuitPaper}.

In this case, our dual-band metasurface absorber almost 100 \% absorbed an incident sine wave not only at 1.29 GHz but also at 3.86 GHz, which corresponded to the third harmonic frequency of 1.29 GHz (Fig.\ \ref{fig:dual}b). When the incident waveform changed to a trapezoidal waveform, this absorber showed an absorptance of 95 \% (Fig.\ \ref{fig:dual}c), which exceeded the absorption performance of the conventional metasurface absorber in Fig.\ \ref{fig:conv} as well as the normalised magnitude of the fundamental wave seen in Table \ref{tab:traSpectrum} (also drawn as the black horizontal line in Fig.\ \ref{fig:dual}c). This is more clearly seen in Fig.\ \ref{fig:dual}d, where each frequency component of the incident and reflected waves was calculated by performing a Fourier transform of their time-domain waveforms. According to this result, the magnitudes of the lowest two frequencies ($f_0$ and 3$f_0$) were reduced to almost zero due to the absorption mechanisms of the two different types of periodic unit cells. We note that higher harmonics (i.e., 5$f_0$, 7$f_0$, $\cdots$) maintained their magnitudes, as the dual-band metasurface absorber did not respond to these frequency components. However, these frequencies are also expected to be suppressed by introducing additional unit cells in a similar manner.

\begin{table}[tb]
 \begin{center}
  \caption{Design parameters used for the dual-band metasurface absorber simulated in Fig.\ \ref{fig:dual}. As shown in Fig.\ \ref{fig:coSim}a, $h$ was used as the substrate height.}
  \label{tab:dual}
  \begin{tabular}{lccccccc} \hline \hline
    Parameter & $p$ & $h$ & $g$ & $l$   &    $C_{f1}$ & $R_{f1}$   & $R_{f3}$     \\ \hline 
    Value & 17 & 3 & 1 & 16 & 4.5 & 500 & 800   \\ 
    Unit & mm & mm & mm & mm & pF & $\Omega$ & $\Omega$   \\ \hline \hline
  \end{tabular}
 \end{center}
\end{table}

\begin{figure}[tb]
\centering
\includegraphics[width=\linewidth]{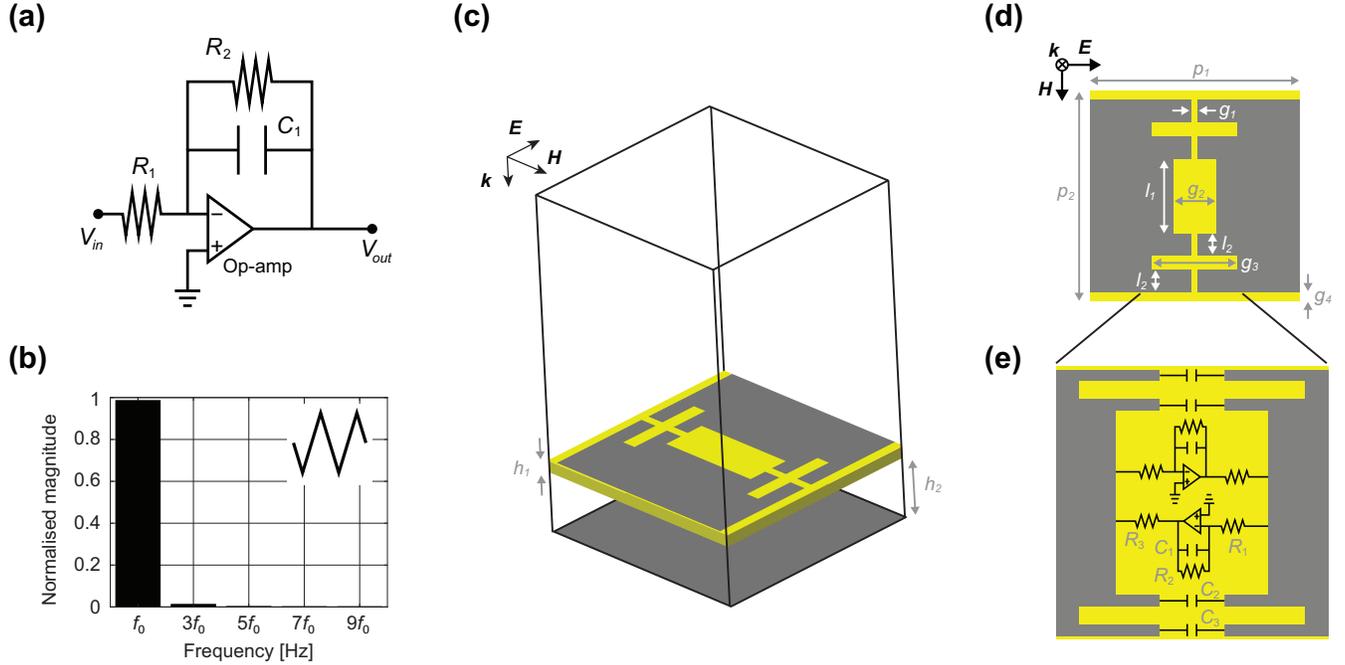}
\caption{(a) Op-amp integrator. (b) Magnitudes of triangular waves at the fundamental frequency and high harmonics. (c, d and e) Unit cell of the waveform-conversion metasurface absorber including a pair of op-amp integrators. The design parameters are given in Table \ref{tab:wc}. }
\label{fig:wcModel}
\end{figure}

For another type of digital noise absorber, we exploited the concept of nonlinear analogue circuits to convert an incident trapezoidal waveform to a different waveform that was more easily absorbed within a metasurface. Specifically, an operational amplifier (op-amp) integrator circuit (Fig.\ \ref{fig:wcModel}a) was used to integrate an input trapezoidal waveform in the time domain to produce a triangular waveform (the inset of Fig.\ \ref{fig:wcModel}b) that had a larger magnitude at the fundamental wave. By following the same procedure as eqs.\ \ref{eq:1stFourier} to \ref{eq:trpG}, the Fourier series and coefficients are obtained by
\begin{eqnarray}
 C_n 
&=& \begin{cases}
\displaystyle \frac{4}{\left( \pi n \right)^2} &\left( n: \rm{odd} \right) \\
    0 &\left( n: \rm{even}\right)
\end{cases},\\
 w\left(t\right) & = & 2\sum^{\infty}_{n=1} \frac{4}{\left( \pi n \right)^2} \sin {2 \pi n f_0 t}.
\end{eqnarray}
Based on the former equation, the normalised magnitude of each frequency component is calculated as seen in Fig.\ \ref{fig:wcModel}b, which analytically shows a larger magnitude at $f_0$ than the one seen in the trapezoidal wave of Fig.\ \ref{fig:waveforms}d. We assumed that if such waveform-conversion circuits were integrated within a metasurface, then more energy would be dissipated at the fundamental wave, which would correspond to the operating frequency of the structure.

Based on this concept, the second type of digital noise absorber was designed as drawn in Figs.\ \ref{fig:wcModel}c, d and e together with the design parameters given in Table \ref{tab:wc}. These figures represent a single unit cell of the structure with paired op-amp integrators and capacitors ($C_{2}$ and $C_{3}$ in Fig.\ \ref{fig:wcModel}e). These additional capacitors were used to adjust the operating frequency of the structure. In addition, the shapes of the conductor edges were not flat (specifically, part of their gaps was closer than others) to realistically design the soldering space for each circuit component. Also, this structure had a vacuum spacer between a substrate (Rogers3003 without loss for simplicity) and a ground place (see $h_2$ in Fig.\ \ref{fig:wcModel}c). Moreover, this absorber used ideal op-amps to simplify the situation and demonstrate the concept of waveform-converting absorption. However, general op-amp integrators still have a frequency dependence and a relationship between the input voltage $V_{in}$ and output voltage $V_{out}$, specifically,
\begin{eqnarray}
 \frac{V_{out} \left( s \right)}{V_{in}\left( s \right)} = -\frac{1}{R_1} \times \frac{1}{\displaystyle\frac{1}{R_2}+sC_1},
\end{eqnarray}
where $s=j\omega$ and $\omega$ is the angular frequency. Provided $1/R_2C_1 \ll \omega$, $V_{out}$ becomes
\begin{eqnarray}
 V_{out}\left(t\right) = -\frac{1}{R_1C_1} \int V_{in} \left(t\right) \: dt,
\end{eqnarray}
which clearly indicates that the input voltage $V_{in}$ is integrated at the output port (i.e., as $V_{out}$).

\begin{table}[tb]
 \begin{center}
  \caption{Design parameters used for the waveform-conversion metasurface absorber simulated in Figs.\ \ref{fig:wcModel} and \ref{fig:wcRes}.  }
  \label{tab:wc}
  \begin{tabular}{lcccccccccccccccc} \hline \hline
    Parameter & $p_1$ & $p_2$ & $h_1$ & $h_2$ & $g_1$ & $g_2$ & $g_3$ & $g_4$  & $l_1$ & $l_2$  & $C_{1}$   & $C_{2}$& $C_{3}$& $R_{1}$   & $R_{2}$ & $R_{3}$     \\ \hline 
    Value & 74.25 & 105 & 1.52 & 50 & 1 & 16 & 54.25 & 3 & 74 & 3 & 3 & 13.22 & 20& 500 & 20 & 730  \\ 
    Unit & mm & mm & mm & mm & mm & mm & mm & mm & mm & mm & pF & pF& pF & $\Omega$ & k$\Omega$ & $\Omega$  \\ \hline \hline
  \end{tabular}
 \end{center}
\end{table}

\begin{figure}[tb]
\centering
\includegraphics[width=\linewidth]{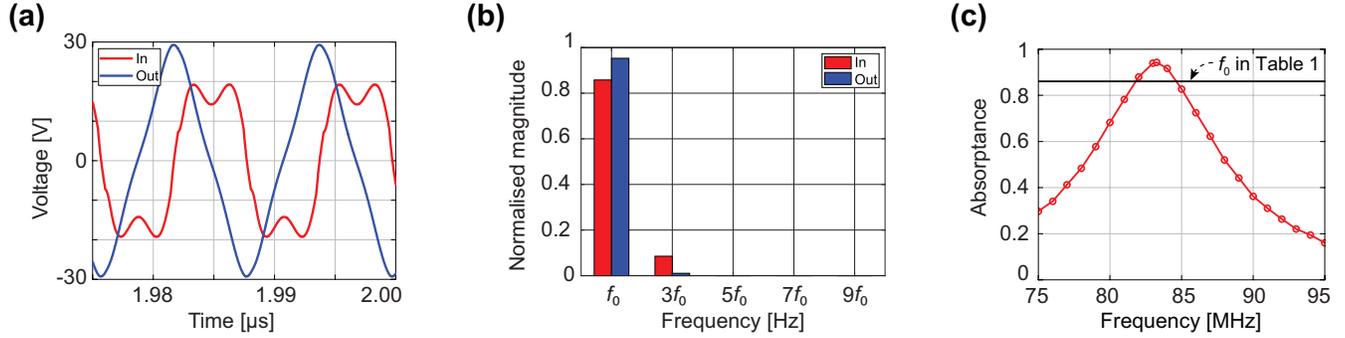}
\caption{(a) Input and output voltages of the op-amp integrator of the waveform-conversion metasurface absorber (with $f_0$ fixed at 83.26 MHz). (b) Magnitudes of the input and output voltages at each frequency component. (c) Absorptance of the waveform-conversion metasurface absorber for the digital waveform as a function of the oscillating frequency $f_0$. }
\label{fig:wcRes}
\end{figure}

The waveform-conversion metasurface absorber shown in Figs.\ \ref{fig:wcModel}c to e was numerically tested as summarised in Fig.\ \ref{fig:wcRes}, where the oscillating frequency of a digital signal was set to 83.26 MHz. First, Fig.\ \ref{fig:wcRes}a plots the input and output voltages at one of the op-amps integrated within the metasurface, which demonstrates that the input waveform was verifiably converted to a different form. Figs.\ \ref{fig:wcRes}a and b indicate that the input waveform differed from an ideal trapezoidal waveform, as the former did not contain frequencies higher than the third harmonic. This is presumably because the metasurface could not respond to these harmonics. However, Fig.\ \ref{fig:wcRes}b shows that the magnitude of the third harmonic was reduced; instead, the fundamental frequency was increased, which ensured that the waveform conversion was successfully performed. In Fig.\ \ref{fig:wcRes}c, the absorption performance of this structure was evaluated using a trapezoidal wave as a function of an oscillating frequency ($f_0$). As a result, this structure exhibited an absorptance of 94 \%, which also exceeded the magnitude of the fundamental frequency of the incident trapezoidal wave (see the black horizontal line of Fig.\ \ref{fig:wcRes}c and $f_0$ of Table \ref{tab:traSpectrum}).

\section*{Discussion}
Our proposed structures were demonstrated to be capable of more efficiently absorbing digital signals than a conventional metasurface. Potentially, conventional broadband absorbers can be also used to absorb digital waveforms \cite{dual1, dualBandAbsTao, My1stAbsPaper}. Although the proposed structures increase absorptance in a limited frequency range, an advantage over broadband absorbers lies in absorbing only the frequency components related to digital noise. In other words, the proposed structures reflect/transmit other signals for wireless communications in adjacent frequency ranges (i.e., our structures minimize the influence on wireless communications). Such signals may not be absorbed by conventional broadband absorbers if their frequency dependence is highly customizable \cite{My1stAbsPaper}. However, we also note that there remain several issues to put the concepts of our absorbers into practice. For example, while our structures were designed to strongly absorb only the lowest two frequency components, other higher harmonics need to be absorbed to further enhance the absorption performance. Additionally, for the sake of simplicity, we fixed the duty cycle of the digital signals used in this study at 50 \%, which means that the information contained in the signals consisted of alternating 1 and 0 values. In reality, however, these values change arbitrarily. Therefore, other frequency components may need to be absorbed with more practical digital signal absorbers. Moreover, experimental validation is necessary to prove the feasibility of these concepts. In particular, since so far dual-band (or even multi-band) absorbers have been experimentally demonstrated by many studies \cite{dual1, dualBandAbsTao, wakatsuchi2012performance}, measurement for the waveform-conversion metasurface absorber would be more important. Here we note that its absorption performance for digital noise is influenced by several additional factors including circuit layout, specific products used as circuit components and their realistic parasitic parameters (especially, those in op-amps). This requires the proposed structure to be further improved but, at the same time, adds more complexity, which makes it difficult to fully clarify the underlying fundamental absorbing mechanism. Therefore, this study only focused on presenting proof-of-concept digital-noise absorbers. Nonetheless, our numerical simulation results show that the proposed structures potentially achieve better absorption performance than conventional metasurface absorbers. Finally, the two digital signal absorbers proposed in this study have both advantages and disadvantages compared to each other. For instance, the dual-band metasurface absorber is easier to design, while it tends to require a physically larger area to fabricate one super cell. This point becomes more important for absorbing the fifth or higher harmonics at the same time in small electronic products. In contrast, the waveform-conversion metasurface absorber is composed of only one type of unit cell, although optimising the structure is not as straightforward as with the dual-band metasurface absorber.

\section*{Conclusion}
We have presented two types of metasurface absorbers to efficiently absorb digital signals. We have shown that digital signals have not only a fundamental frequency but also nonnegligible high harmonics, which limits the absorption performance of conventional metasurface absorbers. One of our proposed absorbers was designed using two kinds of unit cells, each of which absorbed either a fundamental frequency or third harmonic of an incident digital waveform. As a result, this structure was shown to be capable of absorbing 95 \% of the incident energy, which exceeded the magnitude of the fundamental frequency of the incident wave (i.e., approximately 86 \%). The second type of the proposed absorbers exploited the concept of nonlinear analogue circuits to convert an incoming digital signal (in a trapezoidal waveform) to a triangular waveform with a larger magnitude at the fundamental frequency. Simulation results showed that such a waveform-conversion metasurface more strongly absorbed a digital signal than a conventional metasurface absorber. Although there still remain several issues to put these digital signal absorbers into practice, including experimental validation, our study potentially contributes to mitigating EM interference issues caused by digital noise and realising physically smaller, lighter digital signal processing products for the next generation.

%\bibliography{cwEqCircuits}
\providecommand{\noopsort}[1]{}\providecommand{\singleletter}[1]{#1}%

\section*{Acknowledgements}

This work was supported by the Tatematsu Foundation.

\section*{Author contributions statement}

H.W. designed the entire project and conceived of the concepts of the digital-signal metasurface absorbers. R.A. designed the specific metasurface absorbers and performed the numerical simulations. All the authors discussed the results and contributed to writing the manuscript.

\section*{Competing financial interests}
The authors declare no competing interests.

\end{document}